\documentclass[
    ,final            
  ]
  {aipproc}

\layoutstyle{6x9}

\usepackage{epsfig}
\usepackage{graphicx}
\usepackage{psfrag}
\usepackage{subfigure}

\begin{document}

\title{Classical Solutions of SU(3) Yang-Mills Theory and Heavy Quark 
Phenomenology}

\author{O. Oliveira}{
  address={Centro de Física Computacional, Departamento de Física, Universidade de Coimbra, 3004-516 Coimbra, Portugal},
  email={orlando@teor.fis.uc.pt},
}

\begin{abstract}
It is showed that potentials derived from classical solutions of the SU(3)
Yang-Mills theory can provide confining potentials that reproduce the
heavy quarkonium spectrum within the same level of precision as the Cornell
potential.
\end{abstract}

\maketitle



In order to solve the classical Yang-Mills equations of motion, usually
one writes an ansatz that simplifies the Euler-Lagrange equations and, 
hopefully, includes the relevant dynamical degrees of freedom. 
In \cite{Oliveira03} it was proposed a generalized Cho-Faddeev-Niemi-Shabanov 
ansatz for the gluon field, where the gluon is given in terms of two vector 
fields, $\hat{A}_\mu$ and $Y^a_\mu$, and a covariant constant real scalar 
field $n^a$,
\begin{equation}
 A^a_\mu ~ = ~ n^a \, \hat{A}_\mu ~  + ~ 
                      \frac{3}{2g} f_{abc} n^b \partial_\mu n^c ~ + ~
                      Y^a_\mu \, 
 \label{glue}
\end{equation}
with the constraints
\begin{equation}
   D_\mu n^a ~ = ~ 0, \hspace{0.5cm} n^a Y^a_\mu ~ = ~ 0 \, . 
\label{ligacoes}
\end{equation}
In \cite{Oliveira03} it was showed that the above decomposition of 
the gluon field is gauge invariant but not necessarily complete. In the
weak coupling limit, $g \rightarrow 0$, a finite gluon field requires 
either $n = 0$ or $\partial_\mu n = 0$. If $n=0$, the gluon field is reduced
to a vector field in the adjoint representation of SU(3) gauge group. 
For the other case, $\partial_\mu n = 0$, the gluon is writen in terms of the
vector fields $\hat{A}_\mu$ and $Y^a_\mu$ and includes the previous solution
as a particular case. Accordingly, a field such that $n \ne 0$ or
$\partial_\mu n \ne 0$ does not produce a finite gluon field in the weak
coupling limit and, in this sense, can be viewed as a nonperturbative field.
Among this class of fields, the simplest parametrisation for the covariant
scalar field\footnote{From the constraint equation $Dn = 0$ it follows that
$n^2$ is constant. Our choice was $n^2 = 1$. A different value for the norm of
$n$ is equivalent to a rescaling of $\hat{A}$.} is 
$n^a = \delta^{a1} ( - \sin \theta ) + \delta^{a2} ( \cos \theta )$. Then
\begin{equation}
 A^a_\mu ~ = ~ n^a \, \hat{A}_\mu ~  + ~ 
                      \delta^{a3} \frac{1}{g} \partial_\mu \theta ~ + ~
                      \delta^{a8} C_\mu \, ,
 \label{glue1}
\end{equation}
where $C_\mu = Y^8_\mu$. The classical Lagrangian and equations of motion are
independent of $\theta$ and are abelian like in $\hat{A}_\mu$ and $C_\mu$. 
Among the possible nonperturbative gluons given by (\ref{glue}), the simplest 
configuration has $\hat{A} = C = 0$. The coupling to the fermionic fields 
requires only the Gell-Mann matrix $\lambda^3$, decoupling the different 
colour components. This suggests, naively, that such a  field is able to 
produce either confining, non-confining or free particle solutions for the 
quarks.

The classical equations of motion are independent of $\theta$. However, a 
choice of a gauge condition, provides an equation for this field. For the 
Landau gauge, $\theta$ verifies a Klein-Gordon equation for a massless scalar
field. Note that there is no boundary condition for $\theta$, i.e.
the usual free particle solutions of the Klein-Gordon equation are not the
only possible ones. Indeed, writing 
$\theta(t, \vec{r}) = T(t) V( \vec{r} )$, then
\begin{eqnarray}
 & \frac{T'' (t)}{T (t)} ~ = ~
  \frac{\nabla^2 V(  \vec{r} )}{V ( \vec{r} )} ~ = ~ \Lambda^2 ~ > ~ 0 ~ , & \\
 &  T(t) ~ = ~ a e^{\Lambda t} ~ + ~ b e^{ - \Lambda t} ~ , & \\
 &  V ( \vec{r} ) ~ = ~ \sum\limits_{l,m} ~ V_l (r) ~ Y_{lm} ( \Omega ) ~ , 
         & \\
 & V_l (r) ~ = ~ \frac{\alpha_l}{\sqrt{z}} I_{l + 1/2} ( z ) ~ +  ~
                \frac{\beta_l}{\sqrt{z}} K_{l + 1/2} ( z ) ~ , &
\end{eqnarray}
where $z = \Lambda r$ and $I_{l + 1/2} ( z )$ and $K_{l + 1/2} ( z )$ are
modified spherical Bessel functions of the $1^{\mbox{st}}$ and $1^{\mbox{rd}}$
kind\footnote{Note that, by definition, the mass scale $\Lambda$ is 
independent of a rescaling of the gluon field.}. 
The lowest multipole solution is
\begin{equation}
  V_0 (r) ~ = ~ A \, \frac{ \sinh ( \Lambda r ) }{r} ~ + ~
                B \, \frac{e^{ - \Lambda r}}{r} \, 
  \label{v0}
\end{equation}
and the associated gluon field is given by
\begin{eqnarray}
  &   A^3_0 ~ = ~
   \Lambda \left( e^{\Lambda t} - b  e^{\Lambda t} \right) ~ V_0 ( r ) \, ,
          & \\
  &  \vec{A}^{\, 3}_0 ~ = ~ -
    \left( e^{\Lambda t} - b  e^{\Lambda t} \right) ~ \nabla V_0 ( r ) \, .
          &
\end{eqnarray}

From the lowest multipole solution one can derive a potential, which maybe
suitable to describe heavy quarkonium. Indeed, assuming that quarks do not 
exchange energy, in the nonrelativistic approximation and leading order in 
$1/m$, the spatial function in $A^3_0$, $V_0 (r)$, can be viewed as a 
nonrelativistic potential\footnote{The potential is $ \sim 1/r$ 
for short distances and goes to infinity for large quark distances.} and one 
can try to solve the associated Schr\"odinger equation. For the potential
(\ref{v0}), the wave function goes to zero faster than an exponential
for large quark distances,
\begin{equation}
  \psi ( \vec{r} ) ~ = ~ \exp \Bigg\{ \frac{-2}{\Lambda} ~
                \sqrt{\frac{2A}{m}} ~
                \exp \left( \frac{\Lambda r}{2} \right) \Bigg\} \, .
\end{equation}

\begin{figure}[t]
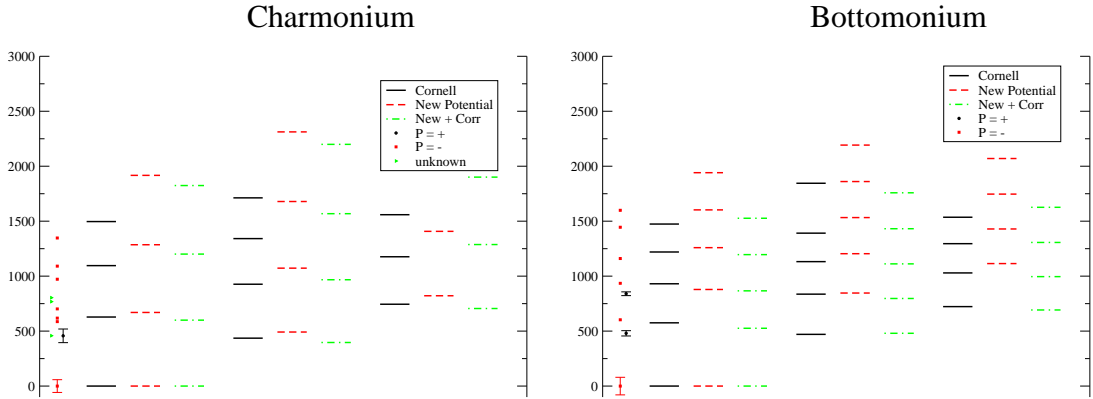

   \begin{minipage}[b]{0.50\textwidth}
      \psfrag{TITULO}{Charmonium}
      \psfrag{SUBTITULO}{}
      \centering
      \includegraphics[origin=c,scale=0.3]{spectrum.cc.Corr.eps}
   \end{minipage}
   \hfill
   \begin{minipage}[b]{0.50\textwidth}
      \psfrag{TITULO}{Bottomonium}
      \psfrag{SUBTITULO}{}
      \centering
      \includegraphics[origin=c,scale=0.3]{spectrum.bb.Corr.eps}
   \end{minipage}
\caption{Heavy quarkonia spectra in MeV. The plots include the spin averaged
experimental values.} \label{prop}
\end{figure}

As a first try to compute the heavy quarkonium spectra, we fixed $A$, $B$
and $\Lambda$ minimising the square of the difference between 
$V_0 (r) + Constant$ and the Cornell potential \cite{Bali} 
$V_{Cornell} = e/r + \sigma r$ ($e = - 0.25$, $\sqrt{\sigma} = 427$ MeV) 
integrated between 0.2 fm and 1 fm. This optimisation provides the following
parameters $A = 5.4$, $B = - 1.0$, $\Lambda = 281$ MeV, 
$Constant = - 1190$ MeV; for these values 
$ -24 MeV \le V_{Cornell} ~ - ~ ( V_0 + Constant ) \le 64$ MeV in the 
integration range considered. Then, we can compare the Schr\"odinger equation 
spectrum for the charmonium ($m_c = 1.25$ GeV) and for the bottomonium 
($m_b = 4.25$ GeV) for the two potentials. The spectrum for the new potential 
shows an equal level spacing for both the charmonium and bottomonium spectra. 
If the $V_0$ charmonium spectrum is quite close to the Cornell spectrum, the 
botomonium shows clear deviations; see figure \ref{prop}. 
The differences are the result of overestimating the strengh of $V_0 (r)$ for
smaller distances. Indeed, one can improve our potential linearising the full
QCD equations around the above configuration. To lowest order, this is
equivalent to add a term like $k/r$ to $V_0$. Computing $k$ 
perturbatively\footnote{For each stationary, the shift in  energy due to this
term is compatible with a perturbative treatment. Corrections are clearly below
10-20\%.} adjusting the $M[(1P)] - M[(1S)]$ bottomonium mass difference, gives
$k = 0.2448251$. The heavy quarkonia spectra, including this correction,
is given in figure \ref{prop}.

In conclusion, classical configurations seem to be able to produce a spectra
close to the Cornell potential. Hopefully, this is an indication that
these configurations can be of help to understand strong interaction physics.
Of course, there are a number of issues that need to be further
investigated (definition of the potential parameters, inclusion of time 
dependence, decay rates). We are currently working on these topics and will
provide a report soon.


\end{document}